\documentclass[prl,nofootinbib,noshowkeys,twocolumn,superscriptaddress]{revtex4}
\usepackage[latin1]{inputenc}
\usepackage{pstricks,pst-node,pst-text,pst-3d,pslatex}
\usepackage[T1]{fontenc}

\usepackage{graphicx}
\usepackage{amsfonts}
\usepackage{amsmath}

\newcommand{\be}{\begin{eqnarray}}
\newcommand{\ee}{\end{eqnarray}}

\begin{document}
\author{P.~Faccioli}
\affiliation{Dipartimento di Fisica  Universit\'a degli Studi di Trento e I.N.F.N, Via Sommarive 14, Povo (Trento), I-38050 Italy.} 
\affiliation{European Centre for Theoretical Studies in Nuclear Physics and Related Areas (E.C.T.$^*$), Strada delle Tabarelle 284, Villazzano (Trento), I-38050 Trento.}
\author{M.~Sega}
\affiliation{C.N.R./I.N.F.M. and Dipartimento di Fisica, Universit\'a degli Studi di Trento, Via Sommarive 14, Povo (Trento), I-38050 Italy.} 
\author{F.~Pederiva}
\affiliation{Dipartimento di Fisica and C.N.R./I.N.F.M.-DEMOCRITOS National Simulation Center,  Universit\'a degli Studi di Trento, Via Sommarive 14, Povo (Trento), I-38050 Italy.}
\author{ H.~Orland}
\affiliation{Service de Physique Th\'eorique,
Centre d'Etudes de Saclay, F-91191, Gif-sur-Yvette Cedex, France.}
\title{Dominant Pathways in Protein Folding}

\begin{abstract}
We present a method to investigate the kinetics of protein folding on a long time-scale and the dynamics underlying the formation of secondary and tertiary structures during the entire reaction. The approach is based on the formal analogy between thermal and quantum diffusion: by writing the solution of the Fokker-Planck equation for the time-evolution of a protein in a viscous heat-bath in terms of a path integral, we derive a Hamilton-Jacobi variational principle from which we are able to compute the most probable pathway of folding.  The method is applied to the folding of the Villin Headpiece Subdomain, in the framework of a Go-model. We have found that, in this model, the transition occurs through an initial collapsing phase driven by the starting coil configuration and a later rearrangement phase, in which secondary structures are formed and all computed paths display strong similarities. This method is completely general, does not require the prior knowledge of any reaction coordinate and represents an efficient tool to perfom ab-initio simulations of the entire folding process with available computers. 
\end{abstract}
\maketitle

Understanding the kinetics of protein folding and the dynamical mechanisms involved in the formation of their structures in an all-atom approach
%
involves simulating a statistically significant ensemble of folding trajectories for a system of $\sim~10^4~$ degrees of freedom. 
Unfortunately, the existence of a huge gap between the microscopic time-scale of the rotational degrees of freedom $\sim~10^{-12}$~s  and the macroscopic time scales of the full folding process  $\sim~10^{-6}-10^1$~s makes it extremely computationally challenging to follow the evolution of a typical $\sim 100$-residue protein for a time interval longer than few tens of nanoseconds.

\begin{figure}
\includegraphics[width=0.6\columnwidth]{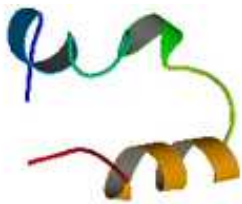}
\caption{Ribbon representation of the Villin Headpiece Subdomain, drawn using Raster3D\cite{ras3d}\label{ribbon}}
\end{figure}

Several approaches have been proposed to overcome such computational difficulties and address the problem of identifying the relevant pathways of the folding reaction~\cite{several}. Unfortunately these methods are either 
affected by uncontrolled systematic errors associated to ad-hoc approximations, or can only be applied to small proteins with typical folding time of the order of few nanoseconds (fast folders).  
In this Letter we present a novel approach to overcome these difficulties: we adopt the Langevin approach and devise a method to {\it rigorously} define and practically compute the {\it most statistically relevant} protein folding pathway.
As a first exploratory application, we have studied the folding transition of the 36-monomer Villin Headpiece Subdomain (PDB code 1VII). This molecule has been extensively studied in 
the literature because it is the smallest polypeptide that has all of the properties of a single domain protein and in addition, it is one of the fastest folders \cite{eaton}.
The ribbon representation of this system is shown in Fig.\ref{ribbon}. 
We analyze the transition from different random self-avoiding coil states to the native state, whose structure was obtained from the Brookhaven Protein Data Bank. 

Our study is based on the analogy between Langevin diffusion and quantum propagation. Previous studies have exploited such a connection to study a variety of diffusive problems using path integral methods~\cite{PI, stonybrook}.
In this work 
we develop the formalism to determine  {\it explicitly} the evolution of the position of {\it each monomer} of the protein, during the entire folding transition, without relying on a specific choice of the reaction coordinate. 

Before entering the details of our calculation it is convenient to review the mathematical framework in a simple case. For this purpose, let us consider Langevin diffusion of a point-particle in one-dimension, subject to an external potential~$U(x)$:
\begin{equation}
\frac{\partial x}{\partial t} = -\frac{D}{k_B T} \frac{\partial U}{\partial x} + \eta(t)
\end{equation}
where $\eta(t)$ is a Gaussian noise with zero average and correlation given by 
$\langle\eta(t)\eta(t')\rangle= 2D\delta(t-t')$. In this equation, $D$ is the
diffusion constant of the particle in the solvent, $k_B$ and $T$ are
respectively the Boltzmann constant and the temperature.

The probability to find the particle at position $x$ at time $t$ obeys the well-known Fokker-Planck Equation:
\be
\frac{  \partial}{\partial\,t}~P(x,t)=D \frac{\partial}{\partial\,x} 
\left(\frac{1}{k_B T}\frac{\partial U(x)}{\partial x}~P(x,t)~\right)
+D\,\frac{\partial^2}{\partial x^2}~P(x,t),\nonumber\\
\label{FPE}
\ee

It is well-known that the stationary solution of (\ref{FPE}) is the Boltzmann distribution $P(x) \sim \exp (-U(x)/k_B T)$.
The solution of  (\ref{FPE}), subject to the boundary conditions 
$x(t_i)=x_i$ and $x(t_f)=x_f$ can be expressed in terms of a path-integral:
\be
\label{path}
P(x_f,t_f|x_i,t_i)=e^{-\frac{U(x_f)-U(x_i)}{2 k_B T}}
\int_{x_i}^{x_f} \mathcal{D}x(\tau)\, 
e^{-{S_{eff}[x]}/2D},
\ee
where $S_{eff}[x]=\int_{t_i}^t d\,\tau~ 
\left(\frac{\dot{x}^2(\tau)}{2}+ V_{eff}[x(\tau)]\right)$,
\be
\label{veff}
V_{eff}(x)=\frac{D^2}{2}\,\left(\frac{1}{k_B T}\frac{\partial
U(x)}{\partial x}\right)^2
-\frac{D^2}{k_B T}\frac{\partial^2~U(x)}{\partial x^2}.
\ee

This result shows that the problem of studying the diffusion of a 
classical particle at temperature $T$ in a medium with diffusion
constant $D$
can be mapped into the problem of determining its 
quantum-mechanical propagation in imaginary time, subject to the 
effective potential $V_{eff}(x)$. 
This approach  has substantial differences from the one introduced in Ref.~\cite{Ons_Mach}, 
where the second derivative of eq.(\ref{veff}) is neglected. 
Such an approximation is not consistent with the Fokker-Planck equation (\ref{FPE}), 
and it leads, at large times, to a distribution which is not the Boltzmann distribution~\cite{ons_mach_wrong}.
 Our approach also differs from the one introduced in Ref.~\cite{elber} where thermal fluctuations were neglected and
friction effects were partially accounted for by choosing large discretization steps to filter-out high-frequency modes.

The most probable path contributing to (\ref{path}) is the one for which the exponential weight $e^{-S_{eff}/2D}$ is maximum, hence for which $S_{eff}$ is minimum. A trajectory which connects configurations that are not classically accessible in the absence of thermal fluctuations  corresponds to an instanton in the quantum-mechanical language. 

The same framework can be applied to study the protein folding, in which the one-instanton solutions represent the most probable folding trajectories
(which we shall refer to as {\it Dominant Folding Pathway}, DFP). 
Determining the DFP for realistic proteins using conventional methods ---such as Molecular Dynamics--- is extremely challenging from the computational point of view. In addition to the  numerical difficulties associated 
with the existence of very different time scales, one has also to face the solution of boundary-value problems, which are considerably harder than  initial-value problems. 

Fortunately, a dramatic simplification is obtained upon observing that the dynamics described by the effective action $S_{eff}$ is energy-conserving and time-reversible. 
This property allows us to switch from the {\it time}-dependent Newtonian description to the {\it energy}-dependent Hamilton-Jacobi (HJ) description. We note that this could not be done at the level of the Langevin equations (or adopting the Onsager-Machlup action). 
In the HJ framework, the Dominant Folding Pathway connecting given initial and final positions is obtained by minimizing --- not just extremizing--- the target function (HJ~functional)
\be
S_{HJ}=\int_{x_i}^{x_f} dl \sqrt{2\left(E_{eff}+V_{eff}[x(l)]\right)}, 
\ee 
where $dl$ is an infinitesimal displacement along the path trajectory. 
$E_{eff}$ is a free parameter which determines the total time elapsed during the transition, according to:
\be
\label{time}
t_f-t_i=\int_{x_i}^{x_f}\,dl \sqrt{\frac{1}{2\left(E_{eff}+V_{eff}[x(l)]\right)}}.
\ee 
It should be stressed that the conserved quantity
$E_{eff}$ 
does not correspond to the {\it physical} energy of the folding transition (which is not conserved in the presence of random forces and friction).
In principle, a  statistical distribution of folding times can be obtained by modeling the statistical distribution of $E_{eff}$ (for example through MD simulations).
In the present work, we adopted the simple choice $E_{eff}=-V_{eff}(x_f)$, which 
corresponds to the longest folding time.
However, we have noted that the minimization of the HJ action by varying the value of $E_{eff}$  of a 
factor up to 5 leads to comparable results. 
\begin{figure}
\includegraphics[width=0.8\columnwidth]{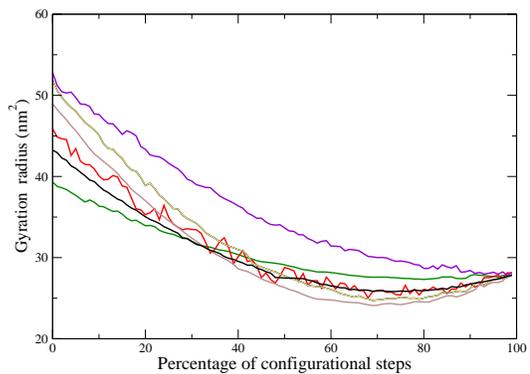}
\caption{The evolution of the radius of gyration as a function of  the
fraction of the total displacement covered during the 
folding transitions in 6 paths corresponding to different initial random coil configurations. 
\label{radius}
}
\end{figure}
The HJ formulation of the dynamics leads to an impressive computational simplification of this problem. In fact, the total Euclidean distance between the coil state and the native state of a typical protein is only 1-2 orders of magnitude larger than the most microscopic length scale, i.e. the typical monomer (or atom) size. As a consequence, only $\sim 100$ discretized displacement steps are sufficient for convergence. This number should be compared with $10^{12}$ time-steps required in the time-dependent Newtonian description. As a result of this simplification, within our approach simulating the entire folding process for a typical protein becomes feasible with available computers.
The physical reason why the HJ formulation is so much more efficient compared to the Newtonian formulation is the following: in traditional Molecular Dynamics simulations, proteins spend most of their time in meta-stable minima, trying  to overcome free-energy barriers. The HJ formulation avoids investing computational times in such "waiting" phases by considering intervals of fixed displacements, rather than fixed time-length. 
 The numerical advantages of the HJ formalism for describing long-time dynamics at {\it constant energy} were first pointed-out in~\cite{elber}.
In this work, we show that comparable computational advantages can also be achieved for {\it stochastic dynamics at fixed temperature}, in which the effects associated with thermal fluctuations and dissipation are consistently taken into account. 
\begin{figure}
\includegraphics[width=0.8\columnwidth]{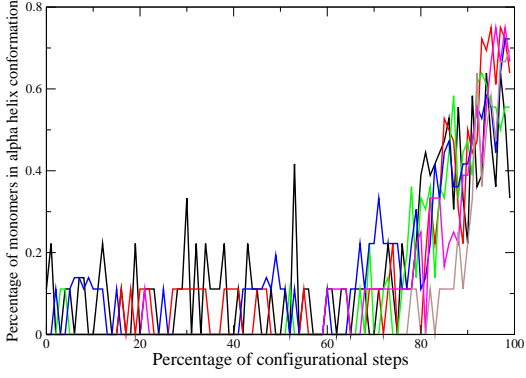}
\caption{The evolution of the percentage of monomers in alpha-helix conformation as a function of  the fraction of the total displacement covered during the folding transitions in 6 paths corresponding to different initial random coil configurations.
\label{alpha}}
\end{figure}

Let us now  apply this formalism to the study of the kinetics of the protein folding.
Although the ultimate goal is to characterize folding pathways using an all-atom description, in this exploratory study we test our method on a very schematic model
in which the effective degrees of freedom (monomers) are representative of amino-acids, and have a fixed mass.
The monomer-monomer interaction is chosen to be the sum of a harmonic bond along the chain, supplemented by a repulsive core between non-consecutive monomers and by an attractive basin between monomers which are in contact in the native state (Go-Model~\cite{gomodel}).  The detailed form of the potential used is:
\begin{eqnarray}
U&=&\sum_{i<j} u({\bf x}_i,{\bf x}_j) 
=\sum_{i<j}\left(\frac{1}{2} K_b (|{\bf x}_i-{\bf x}_{j}|-a\right)^2\delta_{j,i+1} \nonumber\\
&+&\epsilon \sigma_{i,j}\left[ \left(\frac{R_0}{r_{ij}}\right)^12 - \left(\frac{(2\ R_0)^6}{(r_{ij} - R_0)^6 + (2  R0)^6)}\right)\right]\\&+& \epsilon (1-\sigma_{i,j}) \left(\frac{R_{r}}{r_{ij}}\right)^12,\nonumber
\end{eqnarray}
where $r_{ij}=|{\bf x}_i-{\bf x}_{j}|$ and $\sigma_{i j}=1$ if $i$ and $j$ are in native contact, while $\sigma_{i j}=0$ otherwise.
The parameters in the potential have been chosen to be of the same order of similar Go-Model applications~(see \cite{micheletti} and references therein): $a=0.38$~nm, $R_0=0.45$~nm , $R_r=0.65~nm$, $\epsilon=2$~Kcal/mol.
In this first exploratory study we chose to keep the problem as simple as possible and  did not include Coulombic, angular or torsional interactions. 
Hence, the present simple model is not expected to be realistic in predicting the kinetics of tertiary structures formation: the collapse of the protein will be driven mostly by the boundary conditions.
On the other hand, the Go-potential may be sufficiently long-ranged to be effective in the determination of local secondary structures.

The DFP was obtained minimizing numerically the discretized target function 
\be
S_{HJ}=\sum_n^{N-1}\sqrt{2\left(E_{eff}+V_{eff}(n)\right)} \Delta l_{n,n+1}+\lambda P,\ee 
where $P=\sum_i^{N-1} (\Delta l_{i,i+1}-\langle \Delta l\rangle)^2$ and 
\be
V_{eff}(n)&=&\sum_{i} \left[ \frac{D^2}{2 (k_BT)^2} \left(\sum_j {\bf \nabla}_j u({\bf x}_i(n),{\bf x}_j(n))\right)^2 
\right.\nonumber\\&-&\left.\frac{D^2}{k_B T} \sum_j \nabla^2_j u({\bf x}_i(n),{\bf x}_j(n))\right]\\
(\Delta l)^2_{n,n+1}&=&\sum_i({\bf x}_i(n+1)-{\bf x}_i(n))^2,
\ee
$\Delta\,l_{n,n+1}$ is the Euclidean measure of the $n-th$ elementary path step and
$P$ is a penalty function which keeps all the length elements close to their average~\cite{elber} and becomes irrelevant in the continuum limit. 

We have checked that, with 100 discretization steps, simulations performed on a wide range of $\lambda$ lead to consistent results.
\begin{figure}
\includegraphics[width=0.8\columnwidth]{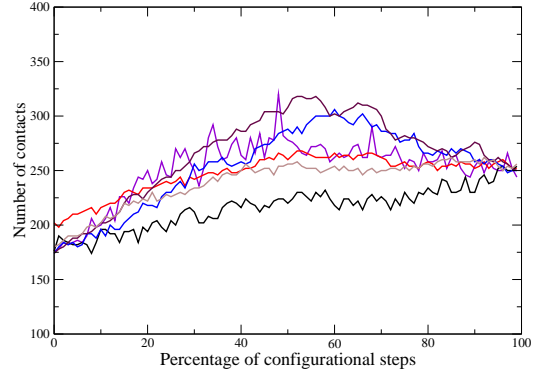}
\caption{The evolution of the number of  contacts as a function of the fraction of the total displacement covered during the folding transitions  in 6 paths corresponding to different initial random coil configurations. \label{contacts}}
\end{figure}
The minimization of the discretized HJ effective action was performed applying an adaptive simulated annealing algorithm and using 50 and 100 path discretization steps. After a preliminary thermalization phase based on usual Metropolis algorithm, we performed about 5 cooling cycles, consisting of 8000 cooling steps each. In order to avoid trapping in local minima, at the begin of each cooling cycle, the configuration was heated-up with few Metropolis steps. At the end of each cooling cycle, the boldness of the Monte Carlo moves was adapted, in order to keep the rejection rate $\sim~90\%$. 
Each calculation lasted for approximately $\sim 12$ hours on a single-processor work station.
We considered the folding transitions from 6 different random self-avoiding coil configurations to the same native state. The center of mass was subtracted form each configurations.  

The results of the simulations performed at $T=300~K$ and damping constant 
$\gamma = \frac{k_B T}{D}=0.1$ns$^{-1}$ are
reported in Fig.~\ref{radius},~Fig.\ref{alpha} 
and  Fig.\ref{contacts} which show respectively the evolution of the
radius of gyration,  the percentage of 
monomers in alpha-helix conformation and the number of 
contacts, as a function of the fraction of the 
total conformational changes. (The total conformational change is
defined as the total Euclidean distance 
covered along the path: $\sum_{n=1}^{N-1}\Delta~l_{n,n+1}$.) 

Some comments on these results are in order. First of all we note 
that, in all simulations performed, the folding transition 
occurs through two rather distinct regimes: in an early stage,  involving the
first $\sim~80\%$ of the total conformational changes, the paths are quite different from each other and no secondary structure is formed. The radius of gyration is decreasing until about $60\%$ of the reaction and then remains essentially constant. Correspondingly, the number of contacts is first increasing and then remains constant.  These results suggest that the initial phase of the folding reaction consits of a collapse of the protein, which strongly depends on the initial coil configuration. 
Only in the last $20\%$ of the conformational evolution, the protein is rearranging to give rise to secondary structures.
This finding is in qualitative agreement with recent experiments on Villin folding kinetics~\cite{eaton} in which the fluorescence quantum yield and frequency shift were investigated with laser temperature-jump. It was found that the unfolding kinetics could be fitted with a bi-exponential function, with time constants of 70~$ns$ and 5 $\mu\,s$. The 70 $ns$ phase was interpreted as related to the formation and melting of the helical turn connecting residues W23 and H27. 

We also note that in the last $20\%$ of the reaction, all paths exihibit some degree of similarity.
This is a natural consequence of the funneled structure of the energy landscape in our topology-based model.


In conclusion, in the present work we have shown how the formal
analogy between  Langevin diffusion and quantum propagation 
can be exploited to perform efficient simulations of the entire protein
folding transition. The framework developed in this work is 
completely general, i.e. it does not rely on the particular choice of
the relevant degrees of freedom nor on the structure of 
the interactions. Unlike other approaches based on a time-dependent description of the dynamics, 
the present approach does not suffer from limitations associated to rare events and therefore its applicability 
is not limited to very small proteins or fast-folders.
A major improvement connected to the use of this approach is the
significant reduction of the computer time necessary for the computation
coming from the different treatment of the fluctuations which determine
the time scale of Newtonian dynamics.
As a result of this simplification, within our approach simulating the entire folding process for a typical protein becomes feasible with available computers.

Since the focus of the present work was on methodology rather than on phenomenology, we have 
performed our exploratory numerical analysis using a coarse-grained topology-based model. We have shown 
that the approach is computationally feasible and allows to access
important information about the evolution of the different 
structures.
We have found that, in such a simple model, the transition occurs through an initial collapsing phase driven by the starting coil configuration and a later rearrangement phase, in which all computed paths display strong similarities.
Simulations using more sophisticated all-atom models are in progress and will clarify whether these are general features or are biases of the topology-based model adopted in this work.

We are indebted to T.~Garel for his help in the initial stage of this work. Special thanks are
due to W.A.~Eaton for important comments and suggestions. We also acknowledge discussions with G.~Colombo, C.~Micheletti, F.~Fogolari and thank L.~Pieri for numerical help. 
Calculations have been partially performed at the E.C.T* supercomputing facility.

\end{document}